\documentclass[paper,12pt]{JHEP3}

\renewcommand{\o}{\omega}

\newcommand{\s}{\sigma}

\renewcommand{\d}{\partial}

\newcommand{\be}{\begin{equation}}
\newcommand{\ee}{\end{equation}}
\newcommand{\bea}{\begin{eqnarray}}
\newcommand{\eea}{\end{eqnarray}}
\newcommand{\bary}{\begin{array}}
\newcommand{\eary}{\end{array}}

\newcommand{\nn} {\nonumber \\}

\newcommand {\eqr} [1]  {{(\ref{#1})}}

\newcommand{\file}[1]{}
\newcommand{\cH}    {{\mathcal{H}}}

\newcommand{\cL}    {{\mathcal{L}}}

\newcommand{\half}  {\frac 1 2}

\usepackage{amssymb}

\preprint{hep-th/0501032\\ BROWN-HET-1436\\ CMS 06004\\ MCGILL-01-05}

\title{Moduli Stabilization  with Long Winding  Strings}

 \author{Yeuk-Kwan E. Cheung $^{1,2}$\footnote{\tt{cheung.edna@gmail.com}},
Scott Watson $^{3}$\footnote{\tt{watson@het.brown.edu}}
and Robert Brandenberger $^{3,4}$\footnote{\tt{rhb@hep.physics.mcgill.ca}}\\
$^1$ Center for Mathematical Sciences\\ 
~~ Zhejiang University,  Hangzhou 310027, China\\
$^2$  Perimeter Institute for Theoretical Physics\\
~~31 Caroline Street North, Waterloo, ON N2L 2Y5, Canada\\
 $^3$ Department of Physics, Brown University, Providence, RI 02912,
U.S.A. \\
$^4$ Department of Physics, McGill University, Montr\'eal, QC,
H3A 2T8, Canada
}

\abstract{%
Stabilizing   all of the modulus  fields  coming  from
compactifications of string theory on internal manifolds is one of the
outstanding challenges for string cosmology.
Here, in a  simple example of  toroidal compactification,
we study the dynamics of the moduli fields corresponding to the size and shape
of the torus along with the ambient flux and long strings winding both internal directions.  
It is known that a  string gas  containing states with non-vanishing winding and momentum 
number in one internal  direction can stabilize the radius of this internal circle to be at self-dual radius.  We show that a  gas of long  strings  winding  all internal  directions can  stabilize  all moduli, except the dilaton which is stabilized by hand,  in this simple example.   
}

\begin{document}
\section{Introduction}

Critical superstring theory \cite{GSW, Pol} is consistent only in
ten space-time dimensions. One possibility to restore consistency
with our observed four-dimensional space-time is to assume that
the six spatial dimensions which are not seen experimentally are
compactified on a manifold with string-scale volume. The size and
shape of the compact internal manifold, along with any flux in the
compactified space, can be parametrized in terms of scalar fields on the observed 
four-dimensional   space-time.
These are the string theoretic moduli fields. In order
to avoid conflicts with observations, there must be a mechanism
which fixes these moduli fields.
In the context of the low energy effective field theory coming
from string theory, it has recently been shown \cite{GKP}
(see also \cite{Acharya}) that the
presence of fluxes can stabilize many but not all of the moduli
fields. In particular, all of the complex structure moduli can
be stabilized, which includes the shape associated with
the internal manifold.  It has, however, proven very difficult to 
stabilize the total volume of the internal manifold using this framework. 

On the other hand,
consider ``string gas cosmology,"  an approach to superstring cosmology
pioneered in \cite{BV} (see also \cite{KP}) and further developed
in \cite{TV}, in which a gas of strings containing, in addition
to the usual effective field theory degrees of freedom,
string winding and momentum modes is coupled to a background
space-time described by dilaton gravity.  It has recently been
shown that this combined action of string winding and momentum
modes can stabilize the radion modulus field at the self-dual radius \cite{Watsona, Subodha}.  
However in~\cite{Watsona, Subodha}  only   ``short'' winding strings were considered, i.e.  string winding in only one internal direction.   The combined effect of momentum modes and winding modes in {\em{each}}  internal direction constrains the radius of this internal circle to be self-dual.  
Thus, it is reasonable to conjecture that it may be possible to stabilize all of the string moduli fields by including fluxes and ``long strings'' winding all internal directions  in the analysis.

In this paper  we consider such a construction in a simple toy model  with moduli
associated with a two dimensional toroidal compactification with flux.  
By populating the torus with a
string gas carrying flux quanta, in addition to momentum and winding charges,
we find that all of the moduli in the problem can be stabilized dynamically with
the exception of the dilaton which we fix by hand in this paper\footnote{The dilaton 
can be likewise trapped  at a particular value by turning on both Neveu-Schwarz and Ramond-Ramond fluxes.  Since we are interested in the stabilization of the complex moduli in this paper, 
we turn on only one kind of flux for simplicity and thus have to  fix the dilaton by hand.}.
We develop a way to incorporate strings winding all internal directions (with the internal momenta consistent with T-duality).  Moduli stabilization is achieved at the classical 
level with extended winding strings and momentum modes alone.  
We analyse the quantum fluctuation of the  system and show that 
both  the shape moduli of the torus as well as  the flux moduli are stabilized dynamically.
This is in contrast to the analyses done in the context of
the low energy field theory action coming from supergravity.  Here we
study the fully time-dependent equations and not just the effective
potential. Dilaton stabilization using tools of string gas
cosmology will be studied in a followup paper.

\section{Background and Toy Model}

Before turning to a review of string gas cosmology and to the
formulation of the toy model which we will study, let us briefly
summarize the status of moduli stabilization in the field theory
limit of string theory, the limit which is most often used as the
starting point for ``string cosmology" (e.g. the papers following
up on \cite{KKLT,KKLMMT} which study the construction of
metastable de Sitter solutions and of inflationary solutions to
the equations of motion). As was realized in \cite{GKP} (see also
earlier work in \cite{previous}), the presence of fluxes can stabilize the
``shape moduli'' of string compactifications. If the internal manifold is a
torus, then the angles of the torus are shape moduli, as are the
ratios of the radii of the individual toroidal directions. A
heuristic argument for the stabilization of the angle modulus is
as follows \cite{Silv}: for fixed values of the two radii of the
torus, the flux energy will be minimal if the volume is largest,
i.e. if the angle between the two cycles of the torus is $\pi /
2$. Similarly, for fixed volume of the torus, the flux energy will
be minimal if the two cycles have identical length. Hence, the
other shape modulus of the torus will also be stabilized by the
flux. On the other hand, the same argument would lead to the total
volume of the torus increasing without bound. Fluxes alone cannot
stabilize the volume modulus. In the original constructions
of \cite{GKP} the dependence on the internal volume drops out of the
potential all together, leading to so-called no-scale models.
This situation was improved in \cite{KKLT}, where non-perturbative corrections
were invoked.  However, as discussed in \cite{sb}, it is not
enough to find a local minimum of the potential, one must also ensure 
that the moduli do
not overshoot such a minimum.  As observed in \cite{kalloshlinde}, 
this remains a challenge
for the KKLT models where the non-perturbative potential is generically 
very shallow.
Progress is being made on the former issue, 
and constructions which also stabilize
the volume modulus have recently been obtained making use of
additional inputs (see e.g. \cite{recent}).  The latter problem of
dynamical stabilization still remains largely neglected.

``String gas cosmology" is an approach to combining string theory
and cosmology which makes crucial use of degrees of freedom and
symmetries which are specific to string theory (as opposed to
point particle field theory). The key degrees of freedom are
string winding modes, and the new symmetry is target space duality
(T-duality). The background space-time is described in terms of
dilaton gravity and matter is taken to be a gas of strings (branes
can also be included \cite{ABE}) containing all degrees of freedom
which are energetically allowed. It is assumed that the background
space contains stable cycles (generalizations were discussed in
\cite{Easther1}). For simplicity, space is often taken to be the
nine-dimensional torus $T^9$. The initial conditions are chosen to
correspond to a hot small universe. Specifically, all spatial
dimensions are of string scale. In the absence of string
interactions, the combination of string momentum and winding modes
keeps all spatial dimensions stabilized at the self-dual radius
($R = 1$ in string units, where $R$ is the radius of each of the
tori). The momentum modes whose energies are quantized in integers
of $1/R$ prevent space from contracting to a singularity while the
winding modes whose energies are quantized in units of $R$ prevent
space from expanding without bound. Thus, string gas cosmology
provides a nonsingular cosmological model \footnote{For further
works on string gas cosmology see \cite{others}. It is important
to note that the conclusions depend crucially on having
fundamental strings or 1-branes in the spectrum of states, and
thus might not extend to the 11-d supergravity corner of the
M-theory space \cite{Easther2} - but see \cite{Stephon}.}.

As argued in \cite{BV}, string intersections will not allow the
disappearance of winding number in more than three spatial
dimensions (in higher dimensions the intersection probability of
string world sheets vanishes), assuming that the net winding
number density vanishes. Numerical support for this argument was
provided in \cite{Sakell}. The evolution of the three large
spatial dimensions in string gas cosmology was studied in detail
in \cite{BEK}, demonstrating that three spatial dimensions can
indeed become large (see \cite{Easther3} for some caveats). Thus,
another major success of string gas cosmology is that it has the
potential to explain why there are only three large spatial
dimensions.

Once the three large spatial dimensions are expanding, the
combined action of the string winding and momentum modes
stabilizes the radii $R_i$ of all other tori to the self-dual
radius \cite{Watsona}. If the value of $R_i$ for some $i = 4, ...,
9$ starts off at a value different from the self-dual radius, it
will perform damped oscillations about the self-dual radius, the
damping term coming from the expansion of the three large
dimensions. The dilaton, however, is not yet stabilized. How to
stabilize the dilaton is in fact one of the major challenges for
string gas cosmology (see e.g. \cite{Berndsen} for a discussion).
A second major challenge is the flatness problem - how to make the
three large spatial dimensions (which begin at a temperature close
to the string scale with string size) sufficiently large to
contain our observed universe. Let us, for the moment, assume that
the dilaton has been stabilized. After the time of stabilization,
the background dynamics is described by the Einstein equations
coupled to the string gas. In this context, it has been shown
\cite{Subodha} that the radii of the extra dimensions which are
still wrapped with winding strings remains stabilized at the
self-dual radius. Crucial to the analysis of \cite{Subodha} is the
inclusion of states with both winding and momentum quantum numbers
which become massless at the self-dual radius (see also
\cite{Watsonb,WatBatt} and in a different context \cite{Kofman}
for a discussion of these states).

These results imply that string gas cosmology has the potential to
stabilize the volume modulus, the one modulus which has proven
problematic to stabilize in the context of the effective field
theory limit of string theory. Since string gases stabilize each
$R_i$ at the self-dual radius, the shape moduli corresponding to
the ratios of $R_i$'s are also automatically stabilized. On the
other hand, string gas cosmology to date has not addressed the
issue of the stabilization of the shape moduli which correspond to
the angles of a torus nor the moduli associated with flux.  
Now if we let the string wind  both directions of the torus,  the potential 
energy of the winding strings will be  minimised  when the torus is a square one.  
Hence we expect extended winding string states will play a role in stabilize the 
angle modulus.  

However, the same heuristic arguments (see \cite{Silv} for a
recent review) which indicate that fluxes can stabilize the shape
moduli of string compactifications also apply in the context of
string gas cosmology. Thus, what we do in this paper is to add
fluxes to the existing framework of string gas cosmology. We study
the simplest toy model in which one angle and flux modulus are free.
Our results show that these moduli are indeed fixed by the string gas carrying
nonzero flux.  Thus, merging string gas cosmology with fluxes appears to 
lead to the {\em dynamical} fixing of all moduli resulting from the string
compactification.

After this brief review of string gas cosmology and why we expect
that by introducing fluxes into the scenario one will be able to
stabilize all of the moduli fields, we will turn to the
formulation of the toy model in which we will study moduli
stabilization. Our starting point is Type II superstring theory on
the background manifold
\be
{\mathcal M} \, = \, {\mathcal R} \times T^9 \, , \ee
where $T^9$ is a nine-dimensional spatial torus. The radii of the
individual toroidal directions are denoted by $R_i$. The
background fields are the space-time metric $G_{MN}$, the
dilaton $\phi$, and the antisymmetric tensor field $B_{MN}$.
These are the fields which are massless in perturbative
superstring theory (see \cite{GSW} for a review). The equations of
motion are the string $\beta$-functional  and will be
discussed in the next section.

The starting point of string gas cosmology is to couple the
background fields to a matter sector consisting of all string
degrees of freedom treated in the ideal gas (i.e. homogeneous)
approximation. Initial conditions are chosen to correspond to an
isotropic string-scale universe, i.e. $R_i = R = 1$ in string
units. Following the arguments of \cite{BV,BEK} we assume that
three of the spatial dimensions become large since in those
dimensions the winding modes can annihilate. Previous work has
shown that the combined action of string winding and momentum
modes will stabilize the other radii at the self-dual radius.

The background contains many moduli fields: the radii $R_i$ of the
individual tori, the angles $\theta_{ij}$ between the i'th and
j'th toroidal direction, the flux on the torus $B_{ij}$,
and the dilaton. As discussed above,
string gas cosmology without fluxes leads to a stabilization of
the overall volume and of the ratio of radii. Thus, the moduli to
focus on are the angles $\theta_{ij}$, the flux $B_{ij}$, and the dilaton.

In the absence of string interactions (intersections), all spatial dimensions
remain small. The self-dual field configuration in this case also
corresponds to a fixed dilaton. Thus, the only moduli left to
worry about are the angles and fluxes. It is sufficient to focus on one
particular angle and flux. Although easily generalizable, for simplicity
we will study compactifications
of the form $\mathbb{R}^{1,3} \times T^2$ with metric
\be
ds^2= -(dx^0)^2+ (d\vec{x})^2+G_{mn} dx^m dx^n,
\ee
where $x^0\equiv t$ is the physical time.  We want to focus on the dynamics
associated with the $T^2$ compactification manifold, so we take 
Minkowski space as the solution for
the ``large'' dimensions which represents a solution of the 
equations of motion ignoring
flux in the large dimensions (this is easily generalized).
It will prove useful to introduce
Greek indices to indicate time along with the compact coordinates,
i.e. $x^{\mu}=(t,x^m)$.  The metric of $T^2$
is parameterized by
\be  \label{metric} G_{m n} \, = \, \left(
        \begin{array}{cc}
           R^2      &   R^2 \sin\theta(t)  \\
           R^2 \sin\theta(t)  &      R^2
         \end{array}
    \right)
\ee
where $\theta$ is the angle of the torus
($\theta = 0$ corresponds to a rectangular torus). Note that we
are using the {{\em dimensionful}} metric for the torus, so the two
coordinates, $x^m$ (which we will later denote by $x$ and $y$), 
run from $0$ to $2\pi$. We turn on flux on the $T^2$ associated with
the antisymmetric tensor field $B_{m n}$ given by
\be \label{B-field} 
B_{mn} \, = \, \left(  \begin{array}{cc}
            0     &   b(t)     \\
            -b(t)   &     0
         \end{array}
    \right)
\ee
In this toy model, the moduli fields to be determined will be the
angle $\theta$ and flux $b(t)$ with the radion moduli fixed by hand
at the value which correspond to the T-dual
symmetry point. We wish to study small fluctuations
about $\theta = 0$ and $b=0$ to demonstrate the stability of this point in
the presence of a string gas with flux.

\section{The Equations of Motion}

Consistency of the string sigma model requires Weyl invariance  at
the quantum level which in turn implies the vanishing of the
$\beta$-functionals of the string fields, namely of the metric
$G_{\mu \nu}$, the rank-two tensor gauge potential $B_{\mu \nu}$,
and the dilaton $\phi$~\cite{Callan:1985ia}.  For a constant
dilaton background they become:
\bea \label{eq:betaG} \beta^G_{\mu \nu} \, &=& \, R_{\mu \nu} +
\frac{1}{4} H_{\mu \kappa \sigma} H_{\nu}^{\; \kappa \sigma} \\
\label{eq:betaB} \beta^{B}_{\mu \nu} \, &=& \, e^{-2\phi} D^\kappa  H_{\kappa
\mu \nu}  \\ \label{eq:betaPhi} \beta^{\phi} \, &=& \,
\frac{D-26}{6\alpha^{\prime}} + R + \frac{1}{12}H_{\kappa \mu \nu}
H^{\kappa \mu \nu}~. \eea
Here, $R_{\mu \nu}$ is  the Ricci tensor, $R$ the Ricci scalar,
$D^{\kappa}$ denotes the covariant derivative, $H_{\mu \kappa
\sigma}$ is the field strength of $B_{\mu \nu}$, $D$ is the number
of space-time dimensions, and $\alpha^{\prime}$ is the string
Regge slope parameter~\footnote{We follow the conventions in
Green, Schwarz and Witten, chapter~3.}. In the presence of matter
the $\beta$-functionals no longer vanish. They are determined by
the matter sources, specifically by the stress-energy tensor
$T^\mu_\nu$ and by the current $J_{\mu \nu}$ of matter.

The Einstein equations in the string frame are obtained by
combining the $\beta$-functions (\ref{eq:betaG}) and
(\ref{eq:betaPhi}) in the following way:
\be  \label{eq:einstein} \beta^G_{\mu \nu} - \half
G_{\mu\nu}\beta^{\phi} \, = \, e^{-2\phi} T_{\mu\nu} \, , \ee
where $T_{\mu \nu}$ is measured in the string frame. Taking the
trace of the Einstein equations one obtains one more equation:
\be  \label{trace} - \half R + \frac{1}{8} H^2 + \frac{3}{2} c =
e^{-2\phi} T^\mu_\mu \ee
which is different from na{\i}vely  setting $\beta^{\phi} = e^{-2\phi}
T^\mu_\mu $.

The flux obeys a Maxwell-like equation given by
\be \label{eq:flux} 
D^\kappa  H_{\kappa \mu \nu} \,  = J_{\mu\nu}
\ee
where the current $J_{\mu \nu}$ is determined by varying the
matter action $S_{matter}$ with respect to $B_{\mu\nu}$
\be \label{eq:Jsource}
J^{\mu\nu} \, = \,\frac{-2}{\sqrt{{-G_D}}} \frac{\d S_{matter}}{\d B_{\mu\nu}} 
\, , \ee
where $G_D$ denotes the determinant of the full space-time metric.
We denote  the components of  $T_{\mu\nu}$ by
\be \label{T} T_{\mu \nu} =  \left(  \begin{array}{ccc}
            ~\epsilon~    &  ~ 0~   &   ~0~   \\
            ~ 0~       &   ~p~    &   ~\tau~    \\
            ~ 0~       &    ~\tau~    &  ~p~
         \end{array}
    \right)
\ee where $\epsilon$ is the energy density and $p$ the pressure
(density). In the presence of a nontrivial angle modulus, we
must add an off-diagonal component $\tau$ in the spatial part of
$T_{\mu \nu}$ for consistency.

Plugging our ansatz for the metric into the Einstein equations
(\ref{eq:einstein}), we obtain component by component the results
\bea \label{eq:tt} 
&{\mathrm{tt:}}& 
 - \frac{1}{4} \dot\theta^2 -\half c + \frac{\dot{b}^2}{4G} =  e^{-2\phi} \epsilon  \\
\label{eq:xx} 
&{\mathrm{xx:}}&   
 -\frac{S_\theta}{C_\theta} \ddot\theta + \frac{1}{4} \dot\theta^2
 + \half  c - \frac{\dot{b}^2}{4G} = \frac{e^{-2\phi}}{R^2} p \\
\label{eq:xy} 
&{\mathrm{xy:}}&
 -\half (\frac{1+S^2_\theta}{S_\theta C_\theta}) \ddot\theta + \frac{1}{4}
\dot\theta^2 + \half  c -\frac{\dot{b}^2}{4G} = \frac{e^{-2\phi}}{R^2S_\theta} \tau 
 \eea
where we have denoted $c \equiv \frac{26-D}{6\alpha^\prime}$. To
shorten the expressions, we have used $S_\theta \equiv
\sin\theta$ and $C_\theta \equiv \cos\theta$. From these
equations  one can derive a consistency condition:
\be  \label{consistency} (1 + S^2_\theta) p - 2\tau S_\theta +
\epsilon R^2 C^2_\theta = 0~. \ee
%

\section{Adding Matter and Fluxes}

We proceed to compute the source terms for the Einstein equations.
We are following the usual approach in string gas cosmology (first
used in \cite{TV}) of treating the string matter sources as an
ideal gas characterized by a homogeneous energy-momentum tensor.
In string gas cosmology it is crucial to consider situations in
which the winding strings fall out of thermal equilibrium
\cite{BV}. Hence, we must use the internal energy of the system
instead of the one-loop free energy as the string-matter source.
We will consider a gas of strings with specified momentum and
winding numbers, plus a homogeneous flux. We will denote the
number density of strings by $\rho$ (not to be confused with an
energy density for which we use the symbol $\epsilon$).

The internal energy, denoted by $E$, can be obtained from one of
the Virasora constraints (see Appendix (\ref{const1})) -
the lack of dynamics for the world sheet metric requires the
world-sheet stress tensor of the string to be identically zero.  
Here we will simply
state the result, with a detailed derivation
given in the appendix. The energy of a string
in the presence of a non-trivial B-field is given by
\bea
E^2&=&\int^{2\pi}_0 d\s \bigl[ G_{mn}p^m p^n +  G_{mn} \acute{X}^m
\acute{X}^n - 2 p^{\,n}  B_{np}   \acute{X}^p
+G^{np} B_{pq} \acute{X}^{ q} B_{nm} \acute{X}^{ m} \bigr] \nn
&&~~~+N +\tilde{N}  -2
\eea
where $m, n = 1, 2$ refer to the compactification manifold $T^2$ and the
integral runs along the length of the
string.  The right- and left-handed oscillatory modes are denoted
$N$ and $\tilde N$, respectively, $p^m=G^{ml}n_l-B^m_l\omega^l$, and
$n_l$ and $\omega^l$ are the momentum and winding numbers
resulting from the compactness of the torus $T^2$.
Explicitly $E^2$ is given by the following expression (ignoring factors of $2\pi$):
\bea  \label{EE}
E^2 &=& G^{11} n_1 n_1 + G^{22} n_2 n_2 + G^{12} n_1n_2 + G^{21} n_2n_1 \nn
	&+&  G_{11} \o_1 \o_1 + G_{12} \o_1\o_2 + G_{21} \o_2\o_1 + G_{22} \o_2\o_2 \nn
	&-& 2n_1G^{11} B_{12}\o_2 - 2 n_1G^{12} B_{21}\o_1
		        - 2n_2 G^{21} B_{12}\o_2 -2 n_2 G^{22} B_{21}\o_1  				\nn
	&+& G^{11}B_{12}\o_2 B_{12}\o_2 + G^{12}B_{12}\o_2B_{21} \o_1   \nn
	&+& G^{21} B_{21}\o_1 B_{12}\o_2 + G^{22} B_{21}\o_1 B_{21}\o_1~.
\eea

In the ideal gas approximation, we take the matter contribution to
the action to be
\be
S_{matter} = \rho \int \sqrt{-G_D} E, \ee
where we recall that $\rho$ is the number density of strings.
From this we obtain the stress-energy tensor $T_{\mu\nu}$, and the
string current $J_{\mu\nu}$:
\bea T_{\mu\nu} \rho^{-1} &=& - E G_{\mu\nu} + \frac{1}{E} \frac{\d
E^2}{\d G^{\mu\nu}} \\ J^{\mu\nu} \rho^{-1} &=&  \frac{1}{E}
\frac{\d E^2}{\d B_{\mu\nu}} 
\eea

The values of the string quantum numbers are constrained by the
second Virasoro constraint (see Appendix (\ref{const1})), namely the 
level matching condition:
\be
n_i \o^i = \tilde{N}- N \, . \ee
Among the states that obey this condition, there are preferred
states, which are massless at the self-dual radius (and in
the absence of flux). These have quantum numbers given by
\be
4 (N - 1) + <n + w, n + w> \, = \, 0 \, , \ee
where $<~,~>$ in the second term indicates  a scalar
product, and $n$ and $w$ are vectors with components $n_i$ and
$w_i$, respectively. These states will dominate the ensemble of
string states if the initial conditions are set up in a
thermal-like state. We will focus on the contribution of states
with $N = 1$, $\tilde{N}=0$ and $n_i = -w_i = \pm 1$.  We take all strings to have
the same momentum and winding quantum numbers in a T-dual ensemble. With equal
probability, we will have any of the following possibilities
\be
\begin{array}{cccc}
           n_1      &   \o_1     &    n_2    &  \o_2      \\
                1       &       -1      &       1       &       -1      \\
                -1      &       1       &       1       &       -1    \\
                1       &       -1      &       -1      &       1       \\
                -1      &       1       &       -1      &       1       \\
\end{array}
\ee
Summing over these states, the average internal energy of the
system becomes:
%
%
\be 
\label{eq:avgE}
 <E^2> =  \frac{n_1^2}{R_1^2 \cos^2 \theta} + \frac{n_2^2}{R_2^2 \cos^2 \theta}+ \o_1^2 R_1^2 + \o_2^2 R_2^2 +\frac{\o_1^2 b^2}{R_1^2 \cos^2 \theta}  + \frac{\o_2^2 b^2}{R_2^2 \cos^2 \theta}  + 2 N~.
\ee
(Note that if we had only kept a subset of these states, we would
have introduced by hand an asymmetry and obtained terms of order
$\theta$ in the expression for the internal energy.)
We have kept the winding and momentum numbers for clarity, even though 
they are actually set to $\pm 1$.  
From this one finds the average contribution of the strings to the
flux,
\bea
\lefteqn{<J_{xy}>=-<J_{yx}> =   \frac{ 1}{E} \cdot \frac{ b(t)}{R^2 \cos^2 \theta}}~,\\
\nn
&&<J_{xx}>=<J_{yy}> =0
\eea

\section{Analysis}

A rectangular torus ($\theta = 0$) with $R = 1$ (self-dual radius)
is a solution of the equations of motion for fixed dilaton and
vanishing flux. We wish to study linear fluctuation around
$\theta=0$ and $b = 0$ and show that the solution is a stable
fixed point. To do this, it is sufficient to expand the expression
for internal energy~(\ref{eq:avgE}) to second order in $\theta$
and to drop all higher order terms. We obtain
\be
E^2 =   \frac{2}{R^2} ( 1 + \theta^2) + 2R^2 + \frac{2b^2}{R^2}
( 1 + \theta^2 + ...) + 2N \ee
In this limit  the expressions for the stress-energy tensor
$T_{\mu\nu}$ and the string source $J_{\mu\nu}$ simplify:
\bea 
T_{xx} &=& T_{yy} = -E + \frac{1}{E} (b^2 +  ( 1+ b^2) \theta^2) \\ 
T_{xy} &=&  - E \sin\theta  + \frac{2}{E} (1 +  b^2) \theta \\ 
\label{eq:J12}    J_{xy} &=&  \frac{b}{E}~. 
\eea
We have set $R=1$, the self-dual radius.


Inserting these results for the energy-momentum tensor into (\ref{eq:xy}) and
(\ref{eq:tt}) we obtain
\bea {\mathrm{xy:}}~ && -\half (\frac{1+ \sin^2\theta}{\sin\theta\,
\cos\theta}) \ddot\theta
        + \frac{1}{4} \dot\theta^2 + \half  c -\frac{\dot{b}^2}{4G}  \nn
        &=& e^{-2\phi} ( - E  +   \frac{2}{E}  ( 1 + b^2) )
\eea
and
\be
{\mathrm{tt:}}~~ - \frac{1}{4} \dot\theta^2 - \half c + \frac{\dot
b^2}{4G} = e^{-2\phi} E \ee
The dependence on $\dot{b}$ and on $c$ can be removed by combining
the two above formulas, yielding the following equation of motion for
$\theta$
\be
\ddot \theta +  4 (1+ b^2) K^{-\half} e^{-2\phi} \theta = 0 \, , 
\ee
where $K\equiv 4+ 2b^2 + 2N$. 
This  is the equation for a stable harmonic oscillator. 
The restoring force receives contributions from 
both the momentum modes and the flux. This is because they both prefer 
the torus to be at  maximum volume, that is at $ \theta = 0$.
The xx-equation and yy-equation  are satisfied at the classical  level and the 
lowest order of perturbations  is  quadratic, which we ignore.

Now let us  turn our attention to the equation of motion for the flux (\eqr{eq:flux})  with source given by \eqr{eq:J12}
\begin{equation} \label{fluxeq}
\ddot b  + \frac{\sin\theta}{\cos \theta} \dot{\theta} \dot{b} \,
= \, - J_{xy}~
\end{equation}
which is satisfied by the classical values of the fields.
We now study the quantum fluctuation.  At linear level the equation becomes:
\begin{equation} \label{fluxeq2}
\ddot b 
+ K^{-\half}  b \, = \, 0 \, .
\end{equation}
In other words the flux also performs harmonic oscillations
about the classical value, $b = 0$. This is consistent with the results of \cite{Ovrut}
and \cite{Copeland} which have studied time-dependent solutions in
the context of the low energy effective field theory actions and
found that the only solution for $b = {\rm const}$ consistent with
our metric ansatz is $b = 0$.

\section{Discussion and Conclusions}

In this paper we studied a simple two-dimensional toroidal background for string gas cosmology.  
We developed a way to incorporate the effect of long strings winding all of the internal directions of the compactification manifold in an attempt to stabilize the angle modulus of the torus.  
We developed a way to  incorporate   long strings  winding all internal directions.  
Moduli stabilization was achieved,  at the classical level,  by these long strings 
 carrying equal winding and momentum charges without resorting to fluxes.
At the quantum level we have shown that in the presence
of fluxes,  the angle between the cycles,  the only shape modulus in
this problem,  is stabilized at a value which maximizes the area given
fixed radii of the torus.  Meanwhile the flux also executes  harmonic oscillations  around its classical value.  It is already known that the combined action
of string winding and momentum modes stabilizes the ratio of the radii
and the total volume.   Hence, we have shown that,  in this example,  all
moduli (except for the dilaton which we have to freeze by hand because there is only Neveu-Schwarz flux in the problem) are stabilized by the long winding strings.    Some other 
study  arrives at the same conclusion by symmetry consideration~\cite{Chatrabhuti}.
Our analysis is based on the solution of the actual dynamical equations of motion
rather that simply by a study of the static effective potential.  

We expect that our main conclusion--namely that in the context of
string gas cosmology all moduli modulo the dilaton can be stabilized
dynamically--extends to more general backgrounds. We also expect
that a more detailed analysis of the action of branes in string
gas cosmology will lead to ways to also stabilize the dilaton.
Work on these topics is in progress.
Phrased differently, our work indicates that the key ingredient
missing in the low energy effective field theory approach to
moduli stabilization is the inclusion of  string winding modes.

\section*{Acknowledgement}

We would like to thank Andrei Linde, Liam McAllister, 
Horatiu Nastase, Subodh Patil, Konstantin Savvidy,
Andrei Starinets  and  Nemani Suryanarayana for useful discussions.
RB is supported in part (at McGill) by an NSERC Discovery grant
and (at Brown) by the US Department of Energy under Contract
DE-FG02-91ER40688, TASK~A.
He thanks the Perimeter Institute for hospitality and
financial support during the period when this project was started.
EC is supported by the visitor programme of Perimeter Institute.
The research at Perimeter is supported by NSERC.
EC thanks the High Energy group of McGill University for hospitality
where the last stage of the project was done.

\section*{Appendix: Energy of String Gas}
\addcontentsline{toc}{section}{Appendix: Energy of String Gas}

Here for completeness we present the standard method for
obtaining the energy of a string gas in the presence of flux 
(see e.g. \cite{Cheung:2003ym}).
The worldsheet action for the string is
\be
        \mathcal{L} = \half \int d^{2} \sigma ~%
        \partial_a X^{m} \, \partial_\beta X^{n}
        \left( \sqrt{h} \, h^{a\beta}  G_{mn} + \epsilon^{a\beta} B_{mn} \right)
\ee
where
\be    \label{app-metricdeterminant}
h= h_{\tau\sigma}\, h_{\tau\sigma} - h_{\tau\tau} \, h_{\sigma\sigma}.
\ee
The generalized momenta are
\be 
\Pi_m = \frac{\delta \mathcal{L} }{\delta \dot{X}^{m} }
        = \sqrt{h} \, G_{mn} (h^{\tau\tau} \dot{X}^{n} + h^{\tau\sigma}  \acute{X}^{ n} ) +
                         B_{mn}  \acute{X}^{ n}
\ee
Now solve for $\dot{X}$
\be
 \dot{X}^{m} =\frac{G^{mn}}{\sqrt{h} \, h^{\tau\tau}} \left( \Pi_n - B_{np} \acute{X}^{ p} \right) -\frac{h^{\tau\sigma}}{h^{\tau\tau}} \, \acute{X}^{ m}
\ee
Eliminating $\dot{X}$ from the action, after a tedious computation,
a simple answer emerges:
\be
\mathcal{L} = \half \int d^{2} \sigma  \, \frac{1}{\sqrt{h} h^{\tau\tau} } \, \left( \Pi^2
            - \acute{X}^{ 2} +  \acute{X}^{ m} B_{mn} G^{np} B_{pq} \acute{X}^{ q} \right).
\ee
Note that we have made use of \eqr{app-metricdeterminant} to eliminate $h^{\sigma\sigma}$.
We then Legendre-transform to arrive at the Hamiltonian
\bea 
\label{app-lightconehamiltonian}
\cH &=&\Pi_m \dot{X}^m - \cL  \nn
        &=& \Pi_m \left( \frac{G^{mn}}{\sqrt{h} h^{\tau\tau}} (\Pi_n -B_{np} \acute{X}^{ p})
                -\frac{h^{\tau\sigma}}{h^{\tau\tau}} \acute{X}^{ m} \right) -\cL  \nn
        &=&  \frac{ 1}{2 \, \sqrt{h} \, h^{\tau\tau} }
        \left[   \Pi^2 + \acute{X}^{ 2}
        - 2 \, \Pi_m G^{mn}B_{np} {\acute{X}}^{ p}
         -  \acute{X}^{ m} B_{mn} G^{np} B_{pq} \acute{X}^{ q}\right]
          - \frac{ h^{\tau\sigma} }{ h^{\tau\tau} } \, \Pi \cdot \acute{X}^{} \nn
\eea

The independent components of the worldsheet metric above play the role
of Lagrangian multipliers  thus a variation with respect to them
gives the contraint equations.
\bea \label{const1}
\Pi^2 + \acute{X}^{ 2} - 2 \,\Pi^m\, B_{mp} \, \acute{X}^{ p}
-\acute{X}^{ m} B_{mn} G^{np} B_{pq} \acute{X}^{ q}= 0 \\
\label{const2}
\Pi \cdot \acute{X} = 0
\eea
The first allows us to obtain the light-cone hamiltonian,
while the second expresses the longitudinal coordinate
in terms of transverse physical degrees of freedom.
Light-cone gauge is selected by setting
$X^+ = 2 \pi a^\prime p^+ \tau$ and
$\Pi_- =  2 \pi a^\prime p^+$.

In this paper we are interested in the effects of a
string gas living on the compact torus $T^2$.  These directions
correspond to the world-sheet fields $X^m$.  The most general solution
respecting our background is given by
\bea
X^{\pm}&=&x^{\pm} + \frac{1}{\sqrt{2}}\,E \tau, \nn
X^2&=&x^2, \;\;\;\;\;\; X^3=x^3, \nn
X^m&=&x^m + \omega^m \sigma + p^m \tau+\mbox{oscillators},\;\;\;\;\;\;\;\;
\eea
where $n_l$ and $\omega^l$ are the momentum and winding numbers
respectively, resulting from the compactness of the torus and we define
$p^m=G^{ml}n_l-2B^m_l\omega^l$.
Plugging this ansatz into the first constraint in (\ref{const1})
 one finds the expression for the energy,
\bea
E^2 &=& \int^{2\pi}_0 d\s \bigl[ G_{mn}p^m p^n +  G_{mn} \acute{X}^m
\acute{X}^n - 2 p^n B_{np}   \acute{X}^p   \nn &&  + G^{np}
B_{pq} \acute{X}^{ q} B_{nm} \acute{X}^{ m} + N +\tilde{N} -2 \bigr],
\eea
or explicitly in our model  $E^2$  is given by the following expression 
(ignoring the unilluminating factors of $2\pi$):
\bea  \label{apEE}
E^2 &=& G^{11} n_1 n_1 + G^{22} n_2 n_2 + G^{12} n_1n_2 + G^{21} n_2n_1 \nn
	&+&  G_{11} \o_1 \o_1 + G_{12} \o_1\o_2 + G_{21} \o_2\o_1 + G_{22} \o_2\o_2 \nn
	&-& 2n_1G^{11} B_{12}\o_2 - 2 n_1G^{12} B_{21}\o_1
		        - 2n_2 G^{21} B_{12}\o_2 -2 n_2 G^{22} B_{21}\o_1  				\nn
	&+& G^{11}B_{12}\o_2 B_{12}\o_2 + G^{12}B_{12}\o_2B_{21} \o_1   \nn
	&+& G^{21} B_{21}\o_1 B_{12}\o_2 + G^{22} B_{21}\o_1 B_{21}\o_1~.
\eea
where we have integrated over the length of the string.

\end {document}